\documentclass[preprint,12pt]{elsarticle}
\RequirePackage{graphicx,epstopdf,adjustbox}
\RequirePackage{tikz,pgfplots}
\usetikzlibrary{shapes,arrows,positioning,matrix,backgrounds,fit}
\RequirePackage[inline]{enumitem}
\RequirePackage{amsmath,amsthm,amssymb,bm}
\RequirePackage{nicefrac}
\RequirePackage{flushend}
\RequirePackage{colortbl}
\usepackage{tcolorbox}
\usepackage{amsmath}
\usepackage{multirow}
\usepackage{url}
\usepackage{lineno}
\setlist{nosep}
\RequirePackage[textsize=scriptsize]{todonotes}
\def\shortname{\textsc{RefOrCite}}

\AtBeginDocument{%
  \providecommand\BibTeX{{%
    \normalfont B\kern-0.5em{\scshape i\kern-0.25em b}\kern-0.8em\TeX}}}

\begin{document}
\begin{frontmatter}

\author{Pradumn Kumar Pandey}
\address{Department of Computer Science and Engineering, IIT Roorkee, India}

\author{Mayank Singh}
\address{Department of Computer Science and Engineering, IIT Gandhinagar, India}

\author{Pawan Goyal}
\address{Department of Computer Science and Engineering, IIT Kharagpur, India}

\author{Animesh Mukherjee}
\address{Department of Computer Science and Engineering, IIT Kharagpur, India}

\author{Soumen Chakrabarti}
\address{Department of Computer Science and Engineering, IIT Bombay, India}

\title{Analysis of Reference and Citation Copying\\ in Evolving Bibliographic Networks}

\begin{abstract}
Extensive literature demonstrates how the copying of references (links) can lead to the emergence of various structural properties (e.g., power-law degree distribution and bipartite cores) in bibliographic and other similar directed networks.  However, it is also well known that the copying process is incapable of mimicking the number of directed triangles in such networks; neither does it have the power to explain the obsolescence of older papers.  In this paper, we propose \shortname, a new model that allows for copying of both the references from (i.e., out-neighbors of) as well as the citations to (i.e., in-neighbors of) an existing node. In contrast, the standard copying model (CP) only copies references.  While retaining its spirit, \shortname{} differs from the Forest Fire (FF) model in ways that makes \shortname{} amenable to mean-field analysis for degree distribution, triangle count, and densification.  Empirically, \shortname{} gives the best overall agreement with observed degree distribution, triangle count, diameter, h-index, and the growth of citations to newer papers. 
\end{abstract}


\end{frontmatter}

\begin{figure}[!tbh]
\centering
\begin{tcolorbox}[colback=gray!25]
\begin{center}
    \textbf{Highlights}
\end{center}
\begin{itemize}
    \item We propose \shortname, a new model that allows for copying of both the references made from (out-neighbors), as well as the citations made to (in-neighbors) a paper.  
    \item We leverage four popular large-scale citation networks to showcase the effectiveness of \shortname.
    \item Empirically and analytically, \shortname, matches the degree distribution better and also generates number of triangles closer to that in real data.
\end{itemize}
\end{tcolorbox}
\end{figure}

\section{Introduction}
\label{sec:intro}

\begin{figure}[th]
\centering
\includegraphics[trim=0 260 220 250, clip, width=0.7\hsize]{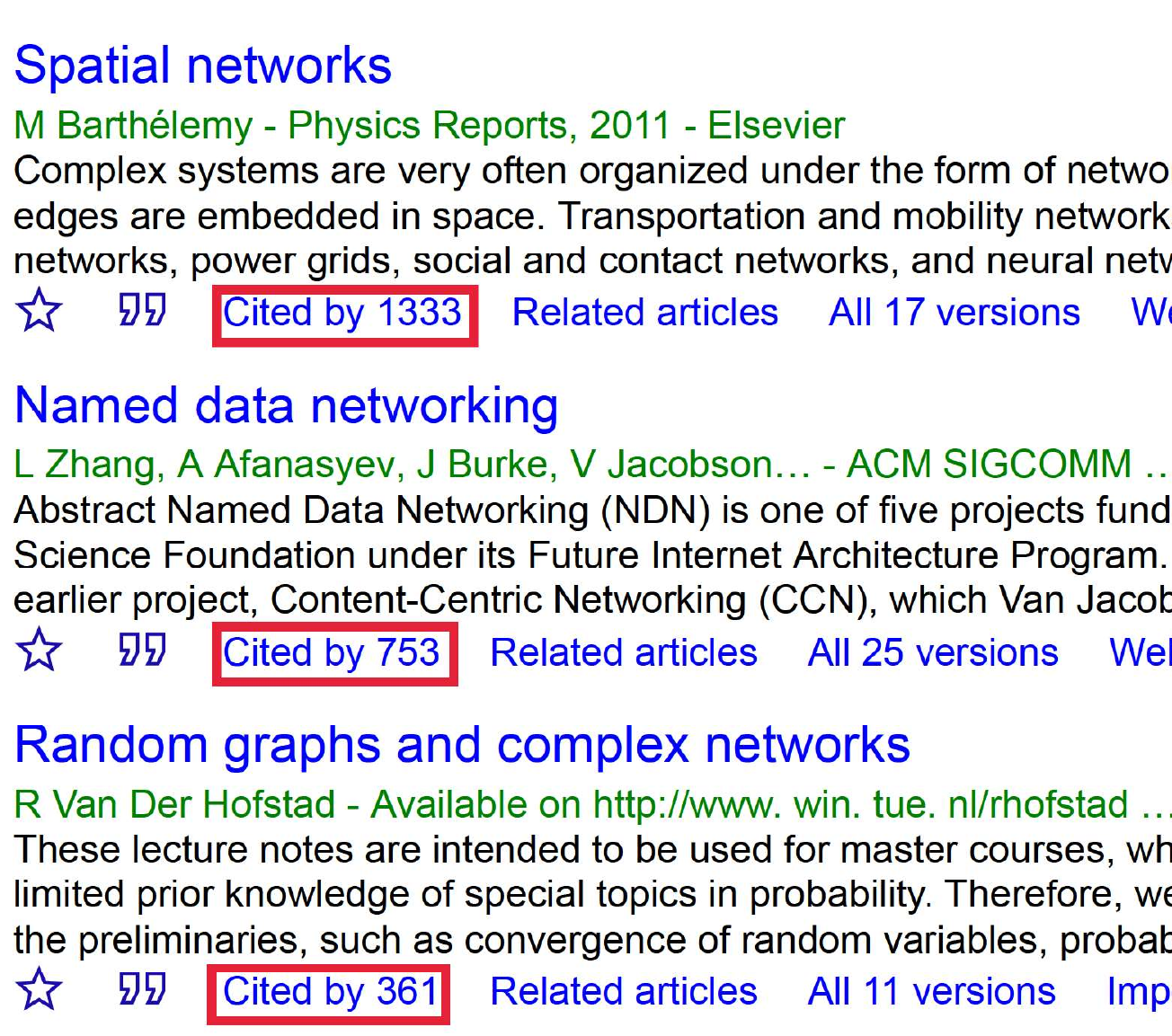}
\caption{Sample Google Scholar result page. Each response article is displayed with a link that leads to papers citing that article.  This feature promotes two ways of copying references: to papers that the article cites, and papers that cite the article.}  \label{fig:citations22}
\end{figure}

Scholarly repositories and retrieval systems, such as Google Scholar (GS), Microsoft Academic Search (MAS), and Semantic Scholar (SS) play a crucial role in scientific information propagation.  Apart from keyword search, they present citing and cited papers, co-author graphs, and related articles.  Online scholarly search is now critical to discovery of related work, thanks to explosive growth of online proceedings and manuscript repositories.  The presentation bias induced by scholarly search can, therefore, significantly influence the evolution of citation networks (see \figurename~\ref{fig:citations22}).

Evolution of citation networks has been under investigation for at least six decades. Many fascinating theories have been proposed to explain the intrinsic forces driving citation.  The motive has been to enable newer models mimic more and more salient properties observed in real networks.  We review a series of standard properties and corresponding modeling approaches in the rest of this section: tail-heavy degree distribution, plentiful bipartite cores and triangles, densification and reduction of diameter as time passes, and the effects of obsolescence of nodes.  In Section~\ref{sec:model}, we propose \shortname, a variation based on the Forest Fire (FF) \cite{leskovec2007graph} and the RelayCite \cite{Singh:2017:RMP:3097983.3098146} models.  The key idea in all three models is that new nodes choose and link to a `base' node, and then nodes in its neighborhood.  As the name suggests, FF explores the neighborhood indefinitely (but limited by a geometric distribution on path lengths).  In contrast, we argue, and later justify experimentally (Section~\ref{sec:eval}), that indefinite exploration is unnecessary in the small-diameter networks seen in practice; it has also made formal analysis infeasible thus far.  In contrast, \shortname{} limits itself to a radius-1 ``controlled burn'', which not only affords formal analysis, but also fits observed networks \emph{better}.

\subsection{Preferential attachment}

The initial set of theories \citep{price1976general, barabasi1999emergence, caldarelli2007scale} focused on the ``rich gets richer effect'', also called the ``Matthew Effect'', based on the premise of preferential attachment (PA).  Although their early popularity was attributed to successful prediction of power-law degree distributions, deviations from power law are well-known.    Going further, Brzezinski~\cite{brzezinski2015power} showed that, in most citation networks, power-law with exponential cut-off and log-normal distributions fit observed data better than pure power law.  Also, if new nodes access the network through a small oligarchy of centralized search engines, degree distribution deviates from power law~\citep{ChakrabartiFV2005Googlearchy}.


\paragraph{Aging and oligarchies:}
PA could not explain obsolescence (loss of popularity over time).  This led to another set of elegant theoretical models of \emph{aging} \citep{Wang20094273, dorogovtsev2002pseudofractal, medo2011temporal, dorogovtsev2013evolution}.  Aging models temper preferential attachment with a temporal decay component, represented by either the log-normal or the exponential distribution.  Fitness parameters, modeling a node's competitiveness to attract links, add another dimension to complex growth models~\citep{bianconi2001competition, wang2013quantifying}. Singh et al.~\cite{Singh:2017:RMP:3097983.3098146} propose an aging model where new node $v$ tentatively chooses a base node $u$, but then cites a node $x$ that cites $u$, in case $u$ is ``too old''.

\subsection{Bipartite cores and the copying model (CP)}

In an effort to preserve power-law degree distribution \emph{and} explain the formation of dense bipartite cores, researchers~\cite{kleinberg1999web, KumarRRSTU2000copy} have proposed a model in which a new node copies all or a fraction of out-degree of the target node (see Figure~\ref{fig:CP}).  At each step, they sample a probability distribution to determine a node $v$ to add edges out of, and a number of edges $k$ that will be added. With probability $\beta$, they add $k$ edges from $v$ to  nodes $V$ chosen independently and uniformly at random. With probability $1-\beta$, they copy $k$ edges from a randomly chosen node to $v$.  Stochastic copying \cite{kumar2000stochastic} is another variant.  Again, at each time step, a new node $u$ enters into the system with $d$ out-links.  To generate the out-links, they begin by choosing a ``prototype'' vertex~$x$. 
The $i^{th}$ out-link  of $u$ is then chosen as follows. With  probability $\alpha$,  the destination  is chosen uniformly  at  random from  $V$, and with the remaining probability the out-link is taken to be the $i^{th}$ out-link of~$x$.  CP exhibits triangle deficiency \cite{ren2012modeling}, cannot accurately model edge-densification \cite{leskovec2007graph} and cannot adequately model aging and obsolescence.

\subsection{Triangle/triad formation}

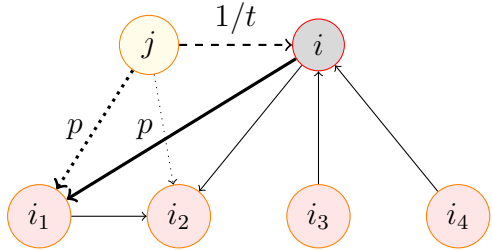
\begin{figure}[th]
\centering
\begin{tikzpicture}
  \node (i) [circle, draw=red, fill=gray!30] {$i$};
  \node (i3) [below=15mm of i, circle, draw=orange, fill=red!10] {$i_3$};
  \node (i2) [left=1cm of i3, , circle, draw=orange, fill=red!10] {$i_2$};
  \node (i1) [left=1cm of i2, circle, draw=orange, fill=red!10] {$i_1$};
  \node (i4) [right=1cm of i3, circle, draw=orange, fill=red!10] {$i_4$};
  \node (j) [left=15mm of i, circle, draw=orange, fill=yellow!10] {$j$};
  \draw[->]  (i) edge [very thick] (i1);
  \draw[->] (i) edge (i2);
  \draw[->] (i1) edge (i2);
  \draw[->] (i3) edge (i);
  \draw[->] (i4) edge (i);
  \draw[->] (j) edge [draw, dashed, thick] node[above] {$1/t$} (i);
  \draw[->] (j) edge [draw, dotted, very thick]  node[left] {$p$} (i1);
  \draw[->] (j) edge [draw, dotted]  node[left] {$p$} (i2);
\end{tikzpicture}
\caption{Copying mechanism of CP model.  A node $j$ newly introduced at time $t$ connects to older base node $i$ with probability $1/t$ and then get connected with one of the first neighbors (out-links only) of node $i$ with probability~$p$.}
\label{fig:CP}
\end{figure}

Triangle formation in real-world networks~\cite{eswaran2018social} is a local event in which a node in the network keeps track of its semi-local structure, for example, neighbours (direct connections) and second-neighbours. It is also more practical due to short visibility of nodes in a large network. In sociology, it is well accepted that the probability of link formation between two nodes would be more if they share more number of common neighbours. There are existing theories and examples which support triangle formation processes in real-world networks. Here, we consider real-world networks which are growing in time but without addition of new edges among older nodes. We do analysis of triangle formation process in citation networks. This is in correspondence to how recommendations of citations by Google Scholar promotes triangle formation process in two ways of copying references: to papers that the article cites, and papers that cites the article. Figure~\ref{fig:CP} shows the CP's copying mechanism.

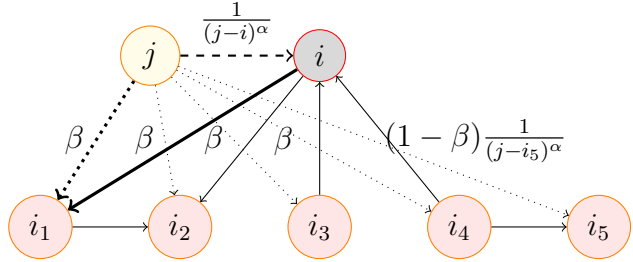
\begin{figure}[th]
\centering
\begin{tikzpicture}
  \node (i) [circle, draw=red, fill=gray!30] {$i$};
  \node (i3) [below=15mm of i, circle, draw=orange, fill=red!10] {$i_3$};
  \node (i2) [left=1cm of i3, , circle, draw=orange, fill=red!10] {$i_2$};
  \node (i1) [left=1cm of i2, circle, draw=orange, fill=red!10] {$i_1$};
  \node (i4) [right=1cm of i3, circle, draw=orange, fill=red!10] {$i_4$};
  \node (i5) [right=1cm of i4, circle, draw=orange, fill=red!10] {$i_5$};
  \node (j) [left=15mm of i, circle, draw=orange, fill=yellow!10] {$j$};
  \draw[->]  (i) edge [very thick] (i1);
  \draw[->] (i) edge (i2);
  \draw[->] (i1) edge (i2);
  \draw[->] (i3) edge (i);
  \draw[->] (i4) edge (i);
  \draw[->] (i4) edge (i5);
  \draw[->] (j) edge [draw, dashed, thick] node[above] {$\frac{1}{(j-i)^{\alpha}}$} (i);
  \draw[->] (j) edge [draw, dotted, very thick]  node[left] {$\beta$} (i1);
  \draw[->] (j) edge [draw, dotted]  node[left] {$\beta$} (i2);
  \draw[->] (j) edge [draw, dotted]  node[left] {$\beta$} (i3);
  \draw[->] (j) edge [draw, dotted]  node[left] {$\beta$} (i4);
  \draw[->] (j) edge [draw, dotted]  node[right] {$(1-\beta)\frac{1}{(j-i_5)^{\alpha}}$} (i5);
\end{tikzpicture}
\caption{Copying mechanism of CPT model.  A node $j$ newly introduced at time $t$ connects to older base node $i$ with probability $\frac{1}{(j-i)^{\alpha}}$ and then get connected with one of the first neighbors of node $i$ with probability~$\beta$.}
\label{fig:CPT}
\end{figure}

\paragraph{Copying with triad formation} ($CPT$):
Krapivsky et al.~\cite{krapivsky2005network} proposed a variant of CP (see Figure~\ref{fig:CPT}).  A newly introduced node randomly selects a target node and links to it, as well as to all out-neighbors (ancestors) of the target node.  Thus, if the target node is the first introduced node (root node), no  additional links are generated by the copying mechanism. If the newly introduced node were to always choose the root node as the target, a star graph would be generated. On the other hand, if the target node is always the most recent one in the network, all previous nodes are ancestors of the target and the copying mechanism would give a complete graph.  In another CPT model \cite{wu2009modeling}, a new node $i$, having out-degree $k_i^{\text{out}}$, selects one of the old node $j$ with an aging probability proportional to its age $t_j = i-j$ to a power $\alpha$ in the existing network as a base node. The rest of the out-degrees of $i$ are attached to random (in- or out-) neighbors of $j$ with probability $\beta$, and otherwise (i.e. with probability $1-\beta$) attach links to older vertices with similar aging probability as above. If there is no available neighbor to attach to then node $i$ selects a new base node as described above and repeats the entire process.  Figure~\ref{fig:CPT} shows the CPT's copying mechanism.

\paragraph{Forest fire} ($FF$) {\it model:}
 Leskovec et al.~\cite[Section\,4.2]{leskovec2007graph} proposed the FF model to remedy some of the above limitations. New node $v$ first chooses a base or `ambassador' node $w$ uniformly at random and links to it.  Then it samples two geometrically distributed numbers $x$ and $y$  with means $p_a/\left(1- p_a\right)$ and $bp_a/\left(1-bp_a\right)$, respectively, where $p_a$ is the forward burning probability and $b$ is the backward burning ratio.  $x$ and $y$  random unvisited in- and out-neighbors of $w$ are visited and linked from $v$.  This step is then recursively applied to the newly linked nodes, but visited nodes are never revisited.  Although FF achieves realistic heavy-tailed degrees, triads, densification and shrinking diameter in experiments, the authors note that ``rigorous analysis of the Forest Fire Model appears to be quite difficult''.  As will become clear, our proposed model, \shortname, resembles FF, but is actually simpler.  This allows a mean-field analysis of degree distribution and expected number of triangles.  Surprisingly, the simplification results in no degradation in the ability to model real networks.  In contrast, aggregating over several graph properties, \shortname{} fits real data better than FF, including the effect of aging, which was not measured for FF earlier.

\subsection{Our contributions}

In this paper, we present a new citation network growth model and call it \shortname.
Like FF and RelayCite \cite{Singh:2017:RMP:3097983.3098146}, \shortname{} is based on real-life scholarly explorations in citation networks.  To trace the origins of an idea, we need to explore back in time, following outlinks.  To discover recent improvements on a result, we need to explore forward in time, traversing inlinks in reverse, facilitated by scholarly search --- see \figurename~\ref{fig:citations22}.

While driven by the same modeling considerations as FF, \shortname{} has important technical differences (see Table~\ref{tab1:growth}).  Like in FF, new node $v$ chooses a base or `ambassador' node $u$, and then expands backward or forward from $u$, but not recursively. 
It helps us derive mean-field estimates for degree distribution, triangle count and some other properties.  We present two variants of \shortname: \shortname1, where, after linking $v$ to $u$, we walk only to newer nodes $x$ that link to $u$ (as in RelayCite \cite{Singh:2017:RMP:3097983.3098146}); and \shortname2, in which, like FF, walking to both in- and out-links of $u$ is allowed.  Accordingly, \shortname{} is characterized by one or two parameters.  Comparing the best-fit forward and backward parameters can reveal key behavioral traits of different scholarly communities.

The majority of previous growth models are evaluated by fitting degree distribution against observed data.  Only a few attempts have been made to match other structural properties like clustering \cite{holme2002growing}, bipartite cores \cite{KumarRRSTU2000copy}, triad formation \cite{xie2015modeling,wu2009modeling}, centrality, or temporal bucket signatures~\cite{Singh:2017:RMP:3097983.3098146}.  It is common for a model that mimics one property well to fit other properties poorly.  A model that provides \emph{simultaneous} good or reasonable fits for many properties can, therefore, be valuable.

Remarkably, the restrictions above that make \shortname{} amenable to analysis do not impair its ability to fit real network properties.  Simulations show that \shortname{} matches observed degree distribution, triangle count, densification, network h-index, and diameter more faithfully than CP and CPT.  Moreover, \shortname{} offers a network-driven explanation of obsolescence, gradually shifting citation focus toward newer papers, in excellent agreement with observed data.


\section{The proposed growth model}
\label{sec:model}

In this section, we present our citation growth model \shortname. 
In \shortname, a new node $j$ appears at time $t+1$ and selects an older `base' node $i$ uniformly randomly from $G_t$, where $G_t$ is the network at time $t$ and $G_{t+1}=\{j\}\cup G_t$.  {The base node is not sampled preferentially: that would increase the selection of older nodes beyond what is warranted by obsolescence models \cite{Singh:2017:RMP:3097983.3098146}.  Uniform base node sampling also leads to heavy-tailed degrees and other realistic properties.}  After introducing the first directed link $(j,i)$ (link is directed towards node $i$), node $j$ may form additional links with immediate in- and out-neighbours of node~$i$. We propose two possible variants on how $j$ forms each such link to the in- and out-neighbours of~$i$:
\textbf{\shortname1}, with a single probability parameter~$p$; and \textbf{\shortname2}, with probabilities~$p_1$ and~$p_2$. The next two  sections describe the formulations of expected growth of the degree of node for our proposed two variants. Table~\ref{tab:growth_equations} compares formulations of expected growth of the degree of node for \shortname and CP. In case of FF and CPT, simple closed form equation is not possible.

\subsection{\shortname1}
Let $\mathcal{N}_i^{\text{in}}(t)$ and $\mathcal{N}_i^{\text{out}}$ be the in-neighbours and out-neighbours of node $i$.  $\mathcal{N}_i(t)=\mathcal{N}_i^{\text{in}}(t) \cup \mathcal{N}_i^{\text{out}}(t)$ denote the set of neighbors of node $i$ at time $t$,  
$\mathcal{N}_i^{\text{in}}(t)$ grows with time, whereas $\mathcal{N}_i^{\text{out}}$ remains fixed after a new node is linked during its introduction into the network. Let $k_i^{\text{out}} = |\mathcal{N}_i^{\text{out}}|$ and $k_i^{\text{in}}(t) =| \mathcal{N}_i^{\text{in}}(t)|$ be out- and in-degree of a node $i$, respectively, with $|\mathcal{N}_i(t)|=k_i(t) = {k}_i^{\text{in}}(t) + {k}_i^{\text{out}}$ being the degree of a node $i$ at time $t$. 
New node $j$ joins the network at time $t+1$.  The expected growth of the degree of node $i$ is given by 
\begin{align}
\dfrac{d k_i(t+1)}{d t} &= \dfrac{1}{t}+\sum_{\mathcal{N}_i(t)}\dfrac{p}{t}, \label{meq}
\end{align}
where $p\in [0,1]$.  The degree of the node $i$ can grow in two ways: either it is selected as the base paper, or it is one of the neighbors of the base paper.  In Eq.~\eqref{meq}, the first term corresponds to selection of the first paper uniformly at random.  At time $t+1$, node $i$ can be the first paper for the new node $j$ with probability~$1/t$.  Also, any neighbour of the node $i$ can be the base paper for node $j$ with the same probability $1/t$, and then $i$ gets a link from $j$ with probability $p$ as a part of the selection of references/citations of the base paper.  This is reflected in the second term.

\subsubsection{Degree Distribution}

In order to obtain an analytical estimate of (expected) degree distribution, we work from Eq.~\eqref{meq}:
\begin{align}
\dfrac{d k_i(t+1)}{d t}&=\dfrac{1+pk_i(t) }{t}  \notag \\
\intertext{By mean field approximation,}
\frac{1}{p} \int \frac{d p k_i(t)}{1 + p k_i(t)} &= \int \frac{dt}{t}  \notag \\
\intertext{Asserting boundary condition $k_i(t_i) = k_i^0$,}
\ln \dfrac{ k_i(t+1) p +1}{k^0_i p+1}&=p \ln \dfrac{t+1}{t_i} \notag \\
\frac{k_i(t+1)+1/p}{k^0_i +1/p}&=\left(\dfrac{t+1}{t_i}\right)^p \notag
\intertext{For $k_i(t)$ to exceed  $k$, we need}
t_i<(t+1)(k+1/p)^{-1/p} & (k^0_i +1/p)^{1/p}.  \notag
\intertext{Since nodes arrive uniformly, we have}
\Pr(k_i>k) & \sim (k+1/p)^{-1/p}(k^0_i +1/p)^{1/p},  \label{prodist}
\end{align}
where $\lim_{t \rightarrow \infty}k_i(t) \rightarrow k_i$.

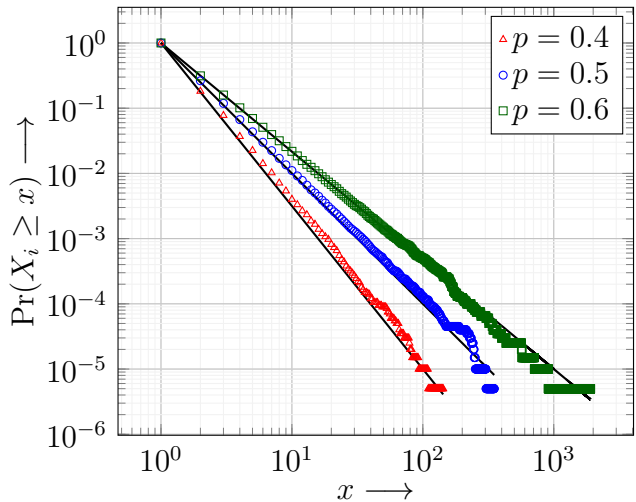
\begin{figure}[ht]
\centering
\begin{tikzpicture}
  \begin{loglogaxis}[
    grid=both, minor grid style={gray!10},
    xlabel={$x\longrightarrow$},
    ylabel={$\Pr(X_i \ge x)\longrightarrow$},
    mark size=1.5pt]
    \addplot [thick, mark=none, forget plot]
    table [x expr=\coordindex+1,y index=0] {fig/xfit0.4.txt};
    \addplot [only marks, color=red, mark=triangle]
    table [x expr=\coordindex+1,y index=1] {fig/xfit0.4.txt};
    \addlegendentry{$p=0.4$};
    \addplot [thick, mark=none, forget plot]
    table [x expr=\coordindex+1,y index=0] {fig/xfit0.5.txt};
    \addplot [only marks, color=blue, mark=o]
    table [x expr=\coordindex+1,y index=1] {fig/xfit0.5.txt};
    \addlegendentry{$p=0.5$};
    \addplot [thick, mark=none, forget plot]
    table [x expr=\coordindex+1,y index=0] {fig/xfit0.6.txt};
    \addplot [only marks, color=green!40!black, mark=square]
    table [x expr=\coordindex+1,y index=1] {fig/xfit0.6.txt};
    \addlegendentry{$p=0.6$};
  \end{loglogaxis}
\end{tikzpicture}
\caption{(Best viewed in color.) Distribution of $X_i =\dfrac{k_i(t+1)+1/p}{k^0_i +1/p}$ is plotted which is proved to be power-law with exponent $1/p$.  Lines correspond to theoretical predictions ($x^{-1/p}$) while markers correspond to simulated results.}  \label{fig:fitting}
\end{figure}

Thus, the degree distribution in \shortname1 closely follows a power-law with a dependency on initial degree (out degree in citation networks).  To work around the initial condition, we consider variable $X_i =\dfrac{k_i(t+1)+1/p}{k^0_i +1/p}$ instead of degree $k_i(t+1)$, and plot $\Pr(X_i>x)$ against $x$ in \figurename~\ref{fig:fitting}.  The event $X_i > x$ corresponds to $(t/t_i)^p > x$, or $t_i < t x^{-1/p}$, implying that $\Pr(X_i > x) \propto x^{-1/p}$, a perfect power law.

\subsubsection{Densification}
Another interesting property of real networks is the behaviour of average connectivity (densification) which we investigate analytically here.  Let $\overline{k}_t$ be the average degree of the network under model Eq.~\eqref{meq} at time $t$.  Then, by counting edges present up to time $t-1$ and adding on $k_t^0$ edges at time $t$, we get
\begin{align}
\overline{k}_{t}&=\dfrac{(t-1) \overline{k}_{t-1} +2k^0_{t}}{t}
\end{align}
The number of edges added at time $t$, i.e., $k_t^0$, can be accounted as one for the link from new node $j$ to the base node, and then $p \overline{k}_{t-1}$ edges to neighbors of the base node.  From this we can approximate
\begin{align}
\overline{k}_{t}&=\dfrac{(t-1)\overline{k}_{t-1}  +2(1+p\overline{k}_{t-1})}{t},\\
\overline{k}_{t}&=\overline{k}_{t-1}
+\dfrac{(2p-1)\overline{k}_{t-1} +2}{t},\\
\dfrac{d \overline{k}_{t-1}}{d t}&=\dfrac{(2p-1)\overline{k}_{t-1} +2}{t}, \\
\intertext{which can be solved as}
(2p-1)\ln \frac{t}{2} &= \ln \dfrac{(2p-1)\overline{k}_{t-1} +2}{2}
\intertext{from which we get}
\overline{k}_{t-1}&=
\begin{cases}
\frac{2}{2p-1}\left(\frac{t}{2}\right)^{2p-1}-2/(2p-1), & p\neq 1/2.\\
2 \ln (t/2)-1, & p= 1/2
\end{cases}
\label{avd}
\end{align}

As the value of $t$ becomes larger

\begin{equation}
    \overline{k}_{t-1}\approx 
\begin{cases}
2/(1-2p), & p < 1/2.\\ \\

\dfrac{2}{2p-1}\left(\dfrac{t}{2}\right)^{2p-1}, & p > 1/2.\\ \\

2 \ln (t/2)-1, & p= 1/2
\end{cases}
\label{phases}
\end{equation}

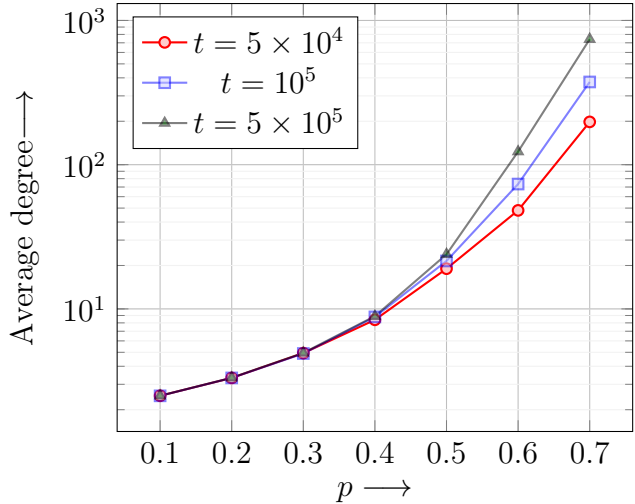
\begin{figure}
\centering
\begin{tikzpicture}
\begin{semilogyaxis}
[ymin=0, legend pos=north west,
    grid=both, minor grid style={gray!10},
    xlabel={$p\longrightarrow$},
    ylabel={Average~degree$\longrightarrow$}]
    \addplot [thick, color=red, mark=*,
      mark options={fill=red!20}]
    table [x index=0,y index=1] {fig/avdphasetrans.txt};
    \addlegendentry{$t=5\times10^4$};
    \addplot [color=blue, thick,
      mark=square*, opacity=0.5,
      mark options={fill=blue!20}]
    table [x index=0,y index=2] {fig/avdphasetrans.txt};
    \addlegendentry{$t=10^5$};
    \addplot [color=black, thick,
      mark=triangle*, opacity=0.5,
      mark options={fill=green!40!black}]
    table [x index=0,y index=3] {fig/avdphasetrans.txt};
    \addlegendentry{$t=5\times10^5$};
\end{semilogyaxis}
\end{tikzpicture}
\caption{(Best viewed in color.) Average degree ($\overline{k}_t$) for different values of~$p$ for networks of different sizes. For $p<0.5$, average degree approaches to a constant value ($\approx\frac{2}{1-2p}$) irrespective of the size of the network and depends on $p$ only, but for $p>0.5$, average degree of the network also depends on the size of the network~($t$). As the size of the network increases, average degree increases. When $p>1/2$, a network of larger size has larger average degree (red circles$<$blue squares$<$black triangles).}  \label{fig:phasetrans}
\end{figure}

Thus, average degree follows power-law in the size of the network, and shows a phase transition around $p=1/2$. In \figurename~\ref{fig:phasetrans}, we plot average degree for different values of $p$ considering networks of  $t=5\times 10^4$ nodes, $t=10^5$ nodes, and $t=5\times 10^5$ nodes. It is observed that for $p<0.5$, average degree converges to same values $\left(\approx\frac{2}{1-2p}\right)$ irrespective of the size $t$ of the networks while for $p>0.5$, average degree of the network also depends on the size of the network. A phase transition in the behaviour of the average degree is observed around $p=0.5$ in \figurename~\ref{fig:phasetrans}.  For $p>1/2$, it shows densification; for $p<1/2$ it asymptotically approaches to a constant average degree $\frac{2}{1-2p}$.

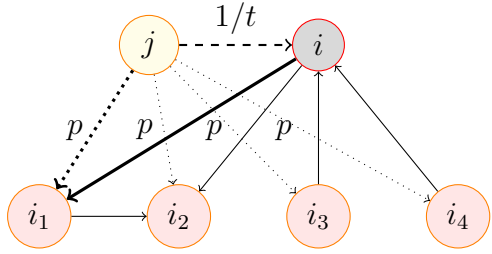
\begin{figure}[th]
\centering
\begin{tikzpicture}
  \node (i) [circle, draw=red, fill=gray!30] {$i$};
  \node (i3) [below=15mm of i, circle, draw=orange, fill=red!10] {$i_3$};
  \node (i2) [left=1cm of i3, , circle, draw=orange, fill=red!10] {$i_2$};
  \node (i1) [left=1cm of i2, circle, draw=orange, fill=red!10] {$i_1$};
  \node (i4) [right=1cm of i3, circle, draw=orange, fill=red!10] {$i_4$};
  \node (j) [left=15mm of i, circle, draw=orange, fill=yellow!10] {$j$};
  \draw[->]  (i) edge [very thick] (i1);
  \draw[->] (i) edge (i2);
  \draw[->] (i1) edge (i2);
  \draw[->] (i3) edge (i);
  \draw[->] (i4) edge (i);
  \draw[->] (j) edge [draw, dashed, thick] node[above] {$1/t$} (i);
  \draw[->] (j) edge [draw, dotted, very thick]  node[left] {$p$} (i1);
  \draw[->] (j) edge [draw, dotted]  node[left] {$p$} (i2);
  \draw[->] (j) edge [draw, dotted]  node[left] {$p$} (i3);
  \draw[->] (j) edge [draw, dotted]  node[left] {$p$} (i4);
\end{tikzpicture}
\caption{Triangle formation.  A node $j$ newly introduced at time $t$ connects to older base node $i$ with probability $1/t$ and then get connected with one of the first neighbors of node $i$ with probability~$p$.  One possible resulting triangle $(j, i, i_1)$ is shown with thicker lines.}
\label{fig:triformation}
\end{figure}

\subsubsection{Triangle formation}
We will now get an analytical estimate of the expected \textbf{number of triangles}. Let $\Delta^i_t$ denote the expected number of triangles attached with node $i$ through time $t$.
Let $\Delta'_{t+1}$ be the expected number of triangles generated at time step $t+1$,
and $\Delta(t+1)$ be the expected total number of triangles in the network generated through time $t+1$.
A new triangle can be generated in two ways:
\begin{enumerate}
\item The new node $j$ gets connected with an older node $i$ with probability $1/t$ and one of its neighbors with probability $p$, for example, triangle $(i,j,i_1)$ in \figurename~\ref{fig:triformation}.
\item During the same growth process new triangles get formed due to existing triangles, for example, triangle $(j,i_1,i_2)$ is formed due to triangle $(i,i_1,i_2)$ in \figurename~\ref{fig:triformation}.
\end{enumerate}
Accordingly, we can write:
\begin{align}
\Delta'_{t+1} = \frac{1}{t} \sum_i p k_i(t) + \frac{1}{t} \sum_i \Delta^i_t p^2 = p\, \overline{k}_t + p^2 \sum_i \frac{\Delta_t^i}{t}
\end{align}
Because each triangle is counted thrice in $\sum_i \Delta^i_t$, we can see that $(1/t) \sum_i \Delta^i_t = 3\Delta(t)/t=3\overline{\Delta}_t$.  Thus we can write
\begin{align}
\Delta'_{t+1} &= p\, \overline{k}_t + 3 p^2 \overline{\Delta}_t\\
\Delta(t+1) &= \sum_{\tau=2}^{t+1} \Delta'_{\tau} \\ 
&= \sum_{\tau=1}^t \left(p\,\overline{k}_\tau
+ 3p^2\overline{\Delta}_\tau\right) = 3p^2 \sum_{\tau=1}^t \overline{\Delta}_\tau + 
p \sum_{\tau=1}^t \overline{k}_\tau \\
\Delta(t+1) &= 3p^2\sum_t\overline{\Delta}_t+ p \sum_t\left(\frac{2}{2p-1}\left(\frac{t}{2}\right)^{2p-1}-\dfrac{2}{2p-1}\right) \\
\Delta(t+1) &=3p^2\sum_t\overline{\Delta}_t+ \dfrac{4p}{2p-1} \left(\frac{t}{2}\right)^{2p}-\left(\dfrac{2p}{2p-1}\right)t.
\end{align}
$\Delta'_{t+1}$, initially, starts with the existence of non-zero value of average degree. This dependency and multiplier $p^2$ results in slower growth of  $3p^2\sum_{\tau=1}^t \overline{\Delta}_\tau$ as compared to $p\sum_{\tau=1}^t \overline{k}_\tau$.  So, for large value of $t$, the triangle count can be approximated in the following way:
 
\begin{equation}
\Delta(t+1)\approx
\begin{cases}
\dfrac{4p}{2p-1}\left(\dfrac{t}{2}\right)^{2p}, & p> 1/2.\\
\dfrac{2p}{1-2p}t, & p< 1/2
\end{cases}
 \end{equation}
 
At $p=0.5$, $\overline{k}_{t-1}=2\ln (t/2)-1$
\begin{align}
\Delta(t+1) &=3p^2\sum_t\overline{\Delta}_t+ p\sum_t (2\ln (t/2)-1).\\
\Delta(t+1) &=3p^2\sum_t\overline{\Delta}_t+ p\left(2\ln t!-2t\ln 2-t\right).\\
\text{Using Stirling's formula}
\nonumber\\
\Delta(t+1) &\approx 3p^2\sum_t\overline{\Delta}_t+ p\left(2t\ln t-2t\ln 2-3t+2\right).\\
\Delta(t+1) &\approx 2pt\ln t.
\end{align}

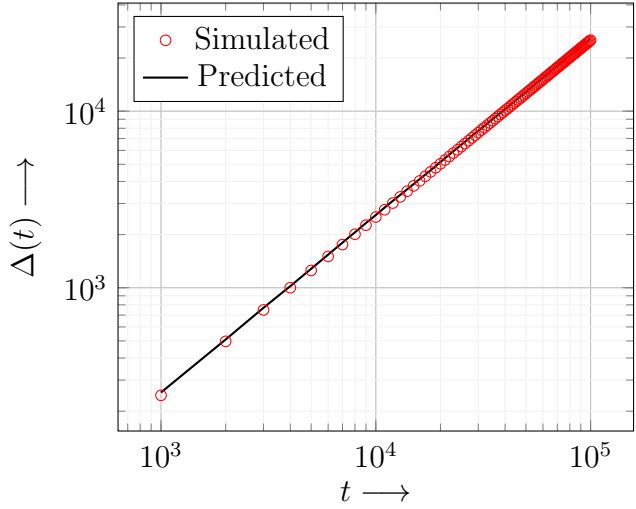
\begin{figure}[ht]
\centering
\begin{tikzpicture}
  \begin{loglogaxis}[legend pos=north west,
    grid=both, minor grid style={gray!10},
    xlabel={$t\longrightarrow$},
    ylabel={$\Delta(t)\longrightarrow$}]
    \addplot [only marks, color=red, mark=o]
    table [x index=0,y index=1] {fig/tri2.txt};
    \addlegendentry{Simulated}
    \addplot [thick, mark=none]
    table [x index=0,y index=2] {fig/tri2.txt};
    \addlegendentry{Predicted}
  \end{loglogaxis}
\end{tikzpicture}
\caption{(Best viewed in color.) Actual number of triangles for the value of $p=0.1$ are plotted in red circles and expected number of triangles are plotted in black line.}
\label{fig:citations1}
\end{figure}

The (analytically obtained) expected number of triangles and the actual number of triangles obtained from a simulated network are shown in \figurename~\ref{fig:citations1}. The theory and the simulation exhibit near-perfect agreement.

\subsection{\shortname2}

In the model \shortname2, copying probabilities of references (out-links) and citations (in-links) are different while in \shortname1, both are same ($p$).
Similar to Eq.~\eqref{meq}, here the expected growth of the degree of node $i$ is given by
\begin{align}
\dfrac{d k_i(t+1)}{d t} &= \dfrac{1}{t}+\sum_{{\mathcal{N}}_i^{\text{in}}(t)}\dfrac{p_2}{t}
+\sum_{{\mathcal{N}}_i^{\text{out}}}\dfrac{p_1}{t},  \label{meq2}
\end{align}
where $p_1, p_2 \in [0,1]$ are the probabilities of copying in- and out-links of the base node.

\subsubsection{Degree distribution}
The procedure followed to compute the degree distribution of \shortname2 is similar to \shortname1. 
\begin{align}
\dfrac{d k_i(t+1)}{d t}&=\dfrac{1+p_2k_i^{\text{in}}(t)+ p_1k_i^{\text{out}}}{t}=\dfrac{F_i^0+p_2k_i(t)}{t},
\end{align}
where $F_i^0 = 1+(p_1-p_2)k_i^{\text{out}}$, and $F_i^0$ is a constant for node $i$ which depends only on the out-degree.
\begin{align}
\Pr(k_i>k) \sim (k+F_i^0/p_2)^{-1/p_2}(k^{\text{out}}_i +F_i^0/p_2)^{1/p_2},
\label{prodist2}
\end{align}
where $\lim_{t\rightarrow \infty} k_i(t)\rightarrow k_i$.

Again, the degree distribution in \shortname2 closely follows a power-law with a dependency on initial degree (out degree in citation networks) with power-law exponent $1/p_2$. 

\subsubsection{Average degree as a function of time} The procedure followed to compute average degree of \shortname2 is similar to \shortname1. Let  $\overline{k}^{\text{in}}_{t-1}$ be the average in-degree  of \shortname2 at time $t-1$. Consider, 
\begin{align}
\overline{k}^{\text{in}}_{t}&=\dfrac{(t-1) \overline{k}^{\text{in}}_{t-1} +k^{\text{out}}_{t}}{t},
\end{align}
which leads to the following result:
\begingroup\allowdisplaybreaks
\begin{align}
\overline{k}^{\text{in}}_{t-1} &=
\begin{cases}
\ln (t/2)-1/2, & p_1+p_2= 1 \\
\left(\frac{1}{p_1+p_2-1} + \frac{1}{2}\right)
\left(\frac{t}{2}\right)^{p_1+p_2-1} \\
\hspace{3em} -\frac{1}{p_1+p_2-1}, & \text{o.w.}
\end{cases}
\label{avd1}
\end{align}
Thus, average in-degree follows power-law in the size of the network, and shows a phase transition around $p_1+p_2=1$. For $p_1+p_2<1$ and $t \rightarrow \infty$, the average in-degree of the networks produced under \shortname2  approach to the fixed point $1/(1-p_1-p_2)$, asymptotically. For $p_1+p_2>1$, \shortname2 shows densification. 

Later, we utilize the relation between model parameters $p_1$ and $p_2$, and average in-degree to simulate the model networks under \shortname2 corresponding to the given real data.
For a given real data, from Eq.~\eqref{avd1}, we evaluate $p_1+p_2$ numerically. Let us say, for a dataset, $p_1+p_2=c$, then during the simulation we select $p_1 \in [0,c]$ (or $p_2$) as free parameter and $p_2=c-p_1$, accordingly.

In \tablename~\ref{t5}, in some cases error for \shortname2 is more as compared to \shortname1 when $p_1+p_2<2p$. The reason is that \shortname2 is not able to explore $p_1=p_2=p$ condition in these cases due to the constraint noted in Eq.~(\ref{avd1}). We point out that if $p_1+p_2 \geq 2p$ then in almost all cases $\text{error}(\shortname2) \le \text{error}(\shortname1)$.

\begin{table*}[!tbh]
\centering
\begin{tabular}{|c| c|} \hline
\textbf{Models} & \textbf{Degree growth formulation}  \\ \hline
FF & Simple closed form equation is not possible \\ \hline
CPT & Simple closed form equation is not possible \\ \hline
 CP & $\dfrac{d k_i(t+1)}{d t} = \dfrac{1}{t}+\sum_{{\mathcal{N}}_i^{\text{out}}}\dfrac{p_1}{t}$\\ \hline
  \shortname1 & $\dfrac{d k_i(t+1)}{d t} = \dfrac{1}{t}+\sum_{\mathcal{N}_i(t)}\dfrac{p}{t}$ \\ \hline
 \shortname2 & $\dfrac{d k_i(t+1)}{d t} = \dfrac{1}{t}+\sum_{{\mathcal{N}}_i^{\text{in}}(t)}\dfrac{p_2}{t}
+\sum_{{\mathcal{N}}_i^{\text{out}}}\dfrac{p_1}{t}$\\ \hline
\end{tabular}
\caption{\label{tab1:growth}Growth equations of different evolution models. Simple closed form equations are not possible in case of FF and CPT.}\label{tab:growth_equations}
\end{table*}

\section{Experimental evaluation}
\label{sec:eval}

\subsection{Datasets}
\label{sec:dataset}

Investigating the questions raised in this work requires rich trajectories of time-stamped network snapshots.  However, such intricately detailed datasets are rare, even while there are an increasing number of new repositories being built and updated regularly\footnote{\url{http://snap.stanford.edu/} is a prominent example.}. We conduct empirical analysis on four citation networks constructed from (i) Biomedical papers\footnote{http://www.ncbi.nlm.nih.gov/pmc/tools/ftp}, (ii) US Supreme Court cases~\cite{fowler2008authority}, (iii) ArXiv's High Energy Physics Theory papers~\cite{snapnets}, and (iv) ArXiv's High Energy Physics - Phenomenology papers~\cite{snapnets}, respectively. As evident from the description, three networks represent scientific article citation networks and one legal document citation network. Biomedical citation network contains papers indexed in PMC Open Access (OA) Subset\footnote{https://www.ncbi.nlm.nih.gov/pmc/tools/openftlist/}. The articles in the OA Subset are made available under a Creative Commons that generally allows more liberal redistribution and reuse than a traditional copyrighted work. The U.S. Supreme Court citation network contains opinions written by the U.S. Supreme Court and the cases they cite from 1754 to 2002 in the United States Reports. The ArXiv citation datasets were originally released as a part of 2003 KDD Cup. It begins within a few months of the inception of the ArXiv, and thus represents essentially the complete history of Physics Theory and Phenomenology papers, respectively. A brief description of each of these networks, along with some bulk statistics, are provided in Table~\ref{t0}. We consider degree distribution, number of triangles, average diameter and obsolescence to compare the real networks against those obtained by the various proposed models. Note that, we keep the same number of nodes in the simulated networks as the corresponding real networks.

\begin{table*}[!tbh]
\centering
\resizebox{\hsize}{!}{
\begin{tabular}{|l| p{120mm}|r|r|} \hline
Networks & Description & Nodes & Edges \\ \hline
Biomedical &   Consists of biomedical papers indexed in NCBI (2001--2008). & 43937 & 162404 
\\ \hline
Supreme court & US Supreme Court cases (1754--2002). Judgements refer to previous judgements.
& 25417 &446490
\\ \hline
ArXiv-HepTH & \textit{High Energy Physics - Theory} papers from arXiv.org (1992--2002).
& 27770 &352807 \\ \hline
ArXiv-HepPH & \textit{High Energy Physics - Phenomenology} papers from arXiv.org (1992--2002).  &  34546& 421578
\\ \hline
\end{tabular}}
\caption{Brief descriptions and salient properties of data sets.}\label{t0}
\end{table*}

\subsection{Degree Distribution}
We adopt the method explained by \cite{pandey2017parametric} to compute the values of parameters under considered models corresponding to each data set. The method is as follows: we discretize $p\in (0,\;1)$ and simulate a model network corresponding to the considered model. $L_1$ distance is computed between degree distribution of the real network and corresponding model network. Value of $p$ is selected corresponding to the minimum $L_1$ distance. Minimized $L_1$ distances are reported in Table~\ref{t5} corresponding to our proposed models (\shortname1 and \shortname2), CP model, CPT model, and FF model.

\begin{table*}[!tbh]
\centering
\resizebox{\hsize}{!}{
\begin{tabular}{|l|c|c|c|c|c|} \hline
Networks &  CP Model&CPT Model& FF Model& \shortname1 & \shortname2 \\ \hline
Biomedical& $1.73$  & \cellcolor{red!15} 3.26 &\cellcolor{green!20}$\textbf{0.42}$ & $0.52$ &$0.75$\\
& $p=0.55$&$\alpha=-1,\; \beta=0.99$ &$b=1,\;p_a=0.001$ & $p=0.41$&$p_1=0.25,\; p_2=0.50$\\ \hline
Supreme court  &$4.67$ & \cellcolor{red!8} 4.18 & $2.37$&$0.95$&\cellcolor{green!20}$\textbf{0.83}$\\
& $p=0.57$&$\alpha=-1,\; \beta=0.99$ & $b=1,\;p_a=0.03$&$p=0.47$&$p_1=0.80,\; p_2=0.17$\\ \hline
ArXiv-HepTH  &$7.84$& \cellcolor{red!15} 8.16 & $3.93$&$1.28$&\cellcolor{green!20}$\textbf{0.65}$\\
& $p=0.58$&$\alpha=-1,\; \beta=0.99$ &$b=10,\;p_a=0.05$ &$p=0.51$ &$p_1=0.40,\; p_2=0.65$\\ \hline
ArXiv-HepPH &$9.18$& \cellcolor{red!8} 7.68 &3.92 &\cellcolor{green!20}$\textbf{0.61}$&\cellcolor{green!20}$\textbf{0.61}$\\
&$p=0.61$ &$\alpha=-1,\; \beta=0.99$ &$b=2,\;p_a=0.04$ &$p=0.52$ &$p_1=0.60,\; p_2=0.43$\\
\hline
\end{tabular}}
\caption{L1 error (smaller is better) between in-degree distributions estimated from simulated networks and corresponding real networks.  Each simulation result is reported for the optimal choice of its parameters, which are also shown in the table.}  \label{t5}
\end{table*}

\begin{figure*}[!tbh]
\def\halfwidth{.47\textwidth}
\centering
\begin{tabular}{cc}
\includegraphics[width=\halfwidth]{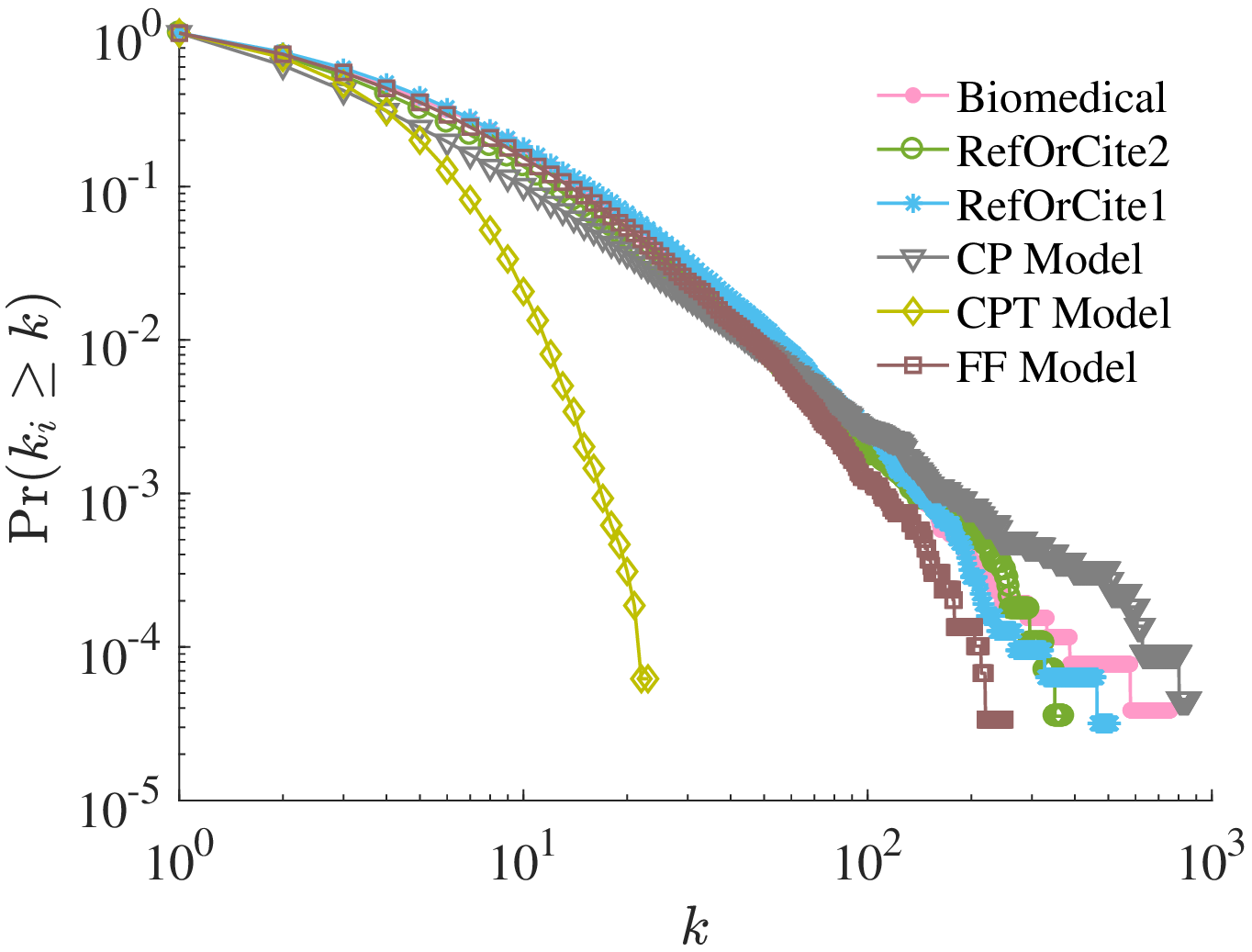} &
\includegraphics[width=\halfwidth]{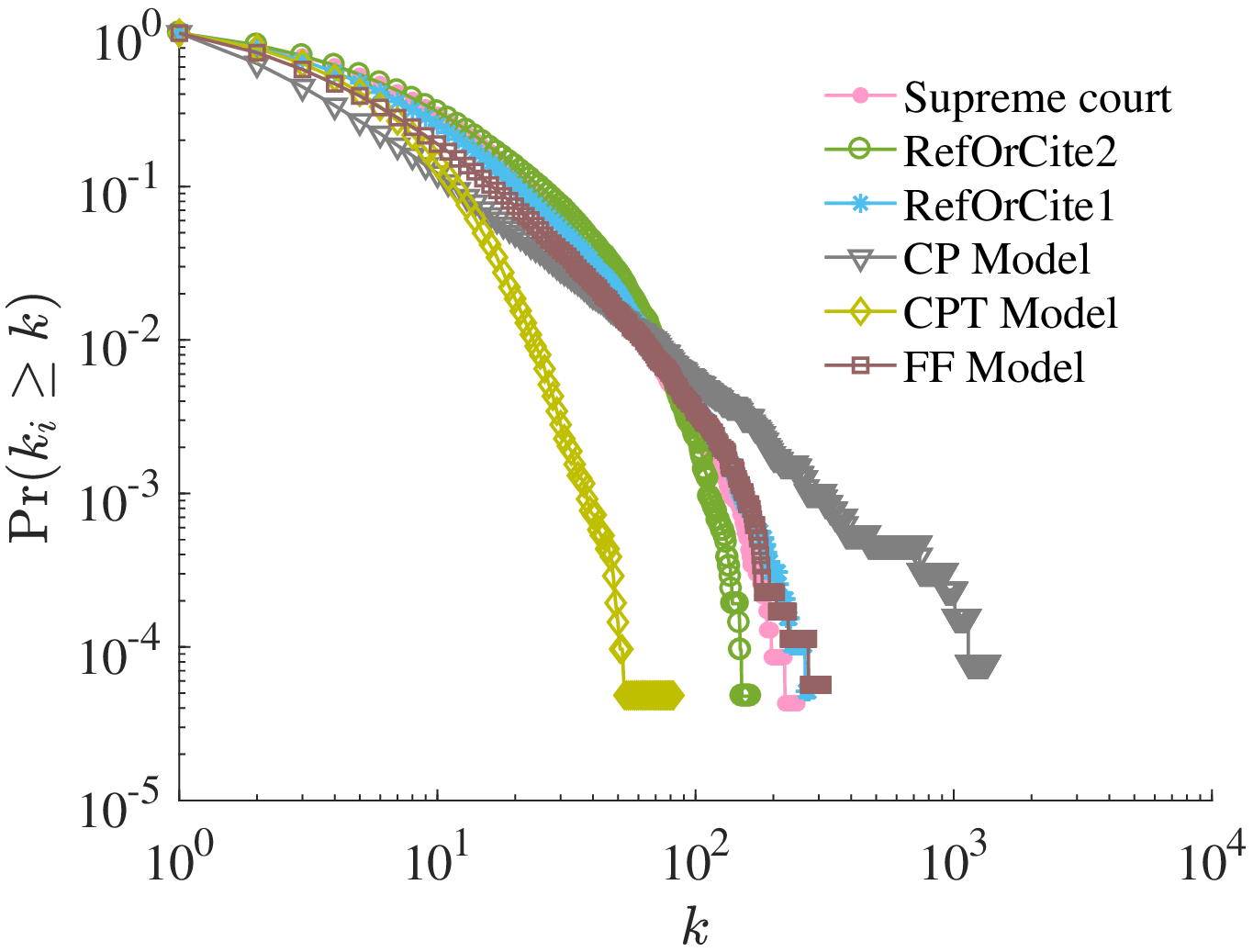}\\
(a) Biomedical  & (b) Supreme court\\
\includegraphics[width=\halfwidth]{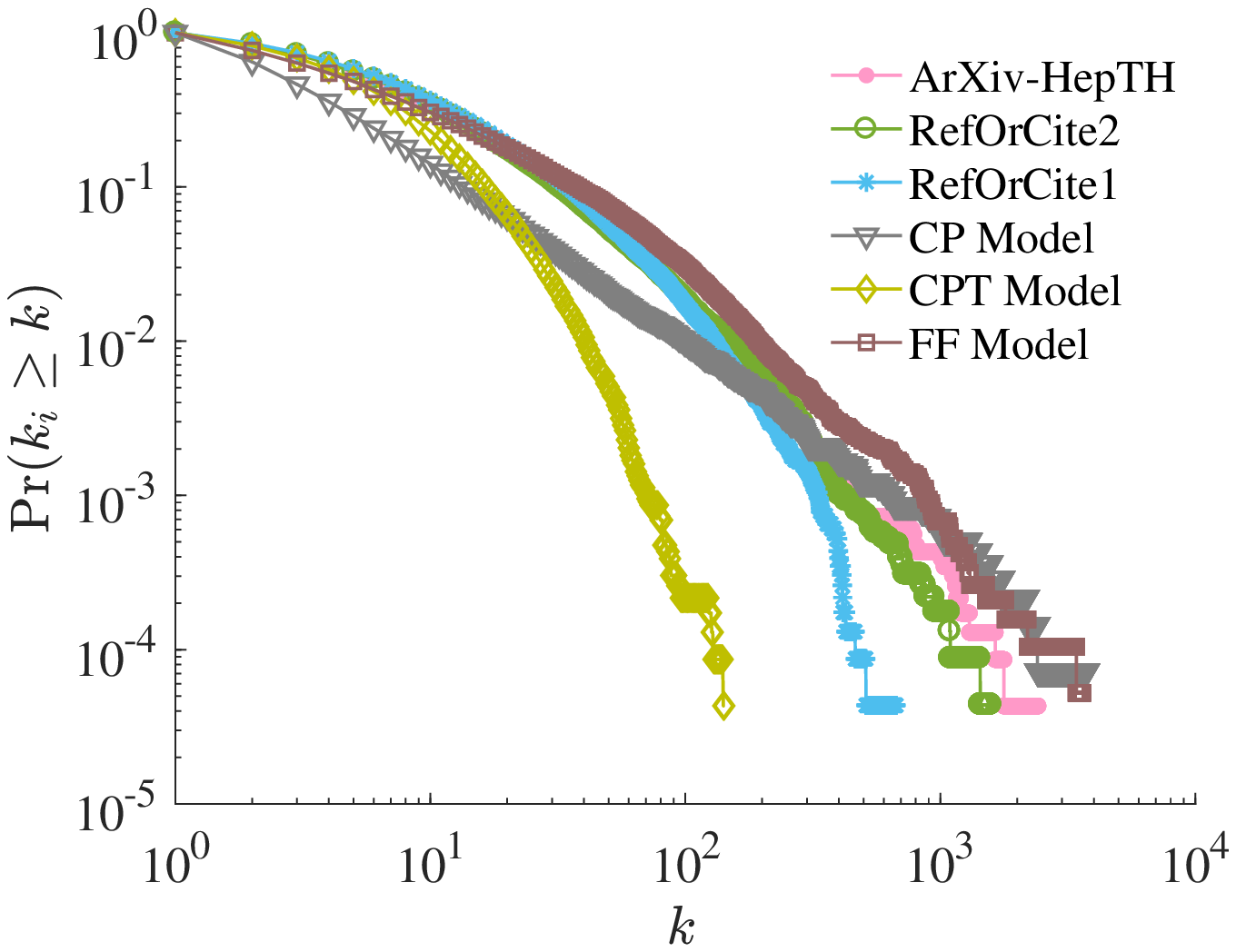} &
\includegraphics[width=\halfwidth]{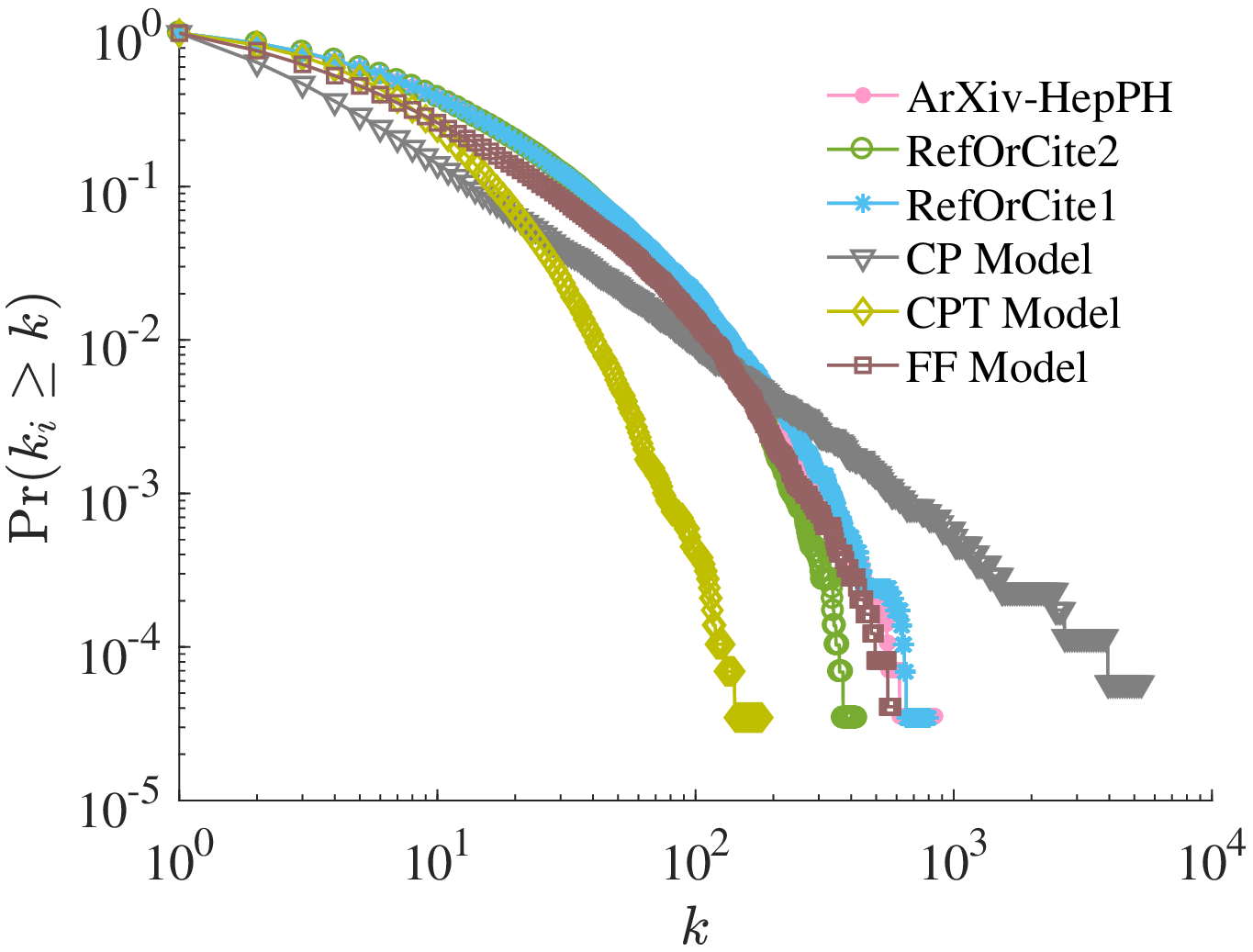}\\
(c) ArXiv-HepTH & (d) ArXiv-HepPH\\
\end{tabular}
\caption{(Best viewed in color.) In-degree distributions for (a)~Biomedical,  (b)~Supreme court, (c)~ArXiv-HepTH, and (d)~ArXiv-HepPH.  Observed data, \shortname2 and \shortname1, CP, CPT, and FF predictions are plotted in pink dots, green circles, cian stars, gray triangles, yellow diamonds, and violet, respectively.}  \label{fig:citations}
\end{figure*}

In \figurename~\ref{fig:citations}, we plot the degree distributions of the four real networks (described in Section~\ref{sec:dataset}).  We compare these with the degree distributions predicted by \shortname1, \shortname2, as well as FF, CPT and CP.  Clearly, \shortname1 and \shortname2 show much better agreement with real data compared to CPT and CP.  The CPT model performs the worse since this model is most suitable for networks that show a slow increase in degree over time; the data sets that we consider on the other hand exhibit faster growth in node degrees. Both \shortname{} variants fit real data as well as the (more complex) FF model.

\begin{figure}[th]
\centering
\includegraphics[trim=0 0 0 0, clip, width=0.65\hsize]{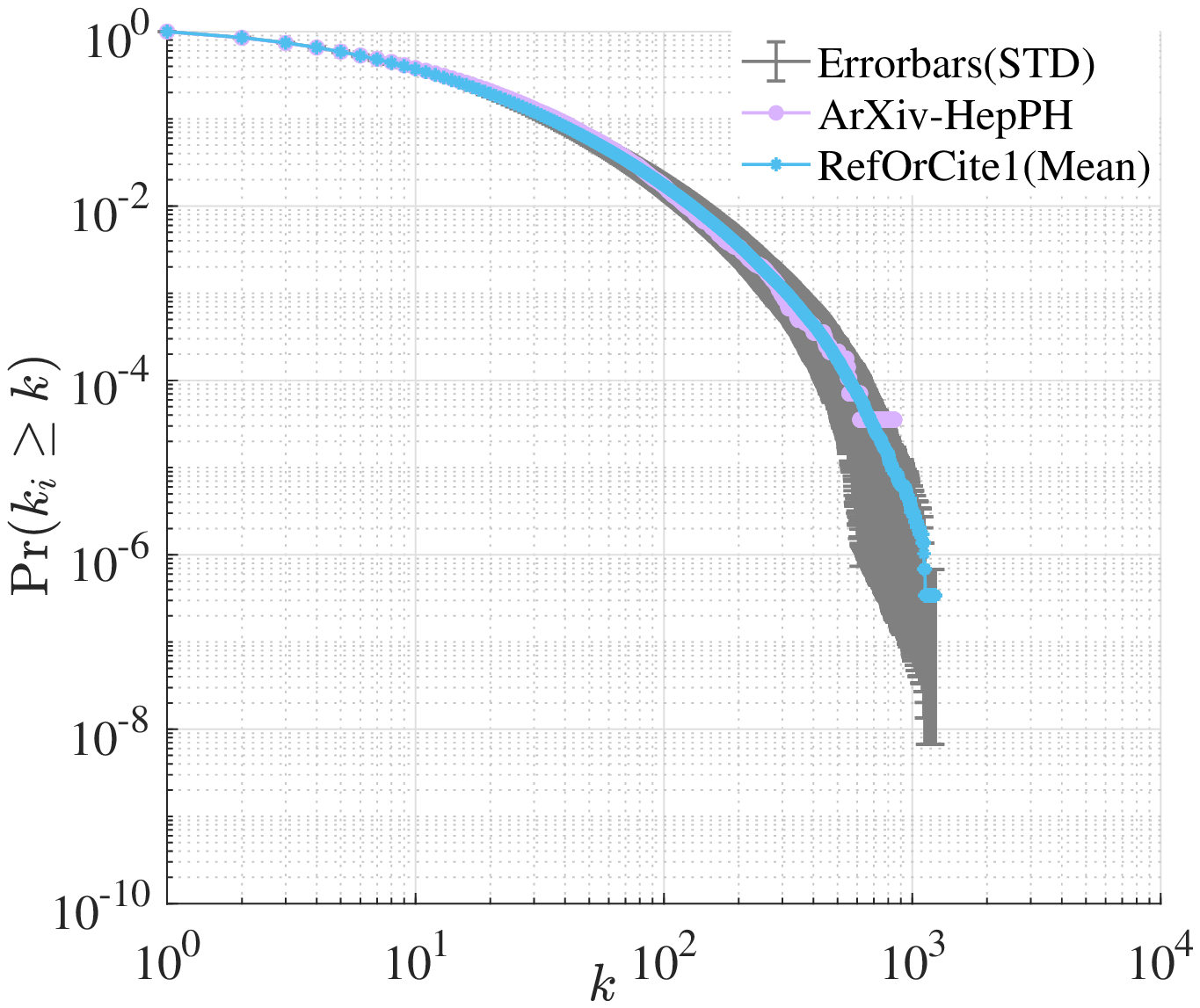}
\caption{Degree distribution of real-network ArXiv-HepPH compared against our model.  Mean of the degree distributions over $100$ model runs obtained under \shortname1{} is plotted in cyan colored stars, and error bars (standard deviations (STD)) are plotted in gray color.  Mean STD (mSTD) over all networks is $0.001$ and mean variance (mVAR) is $7.9\times 10^{-6}$. For same real data set, CP, CPT, and FF have values of (mSTD, mVAR) ($9.9\times 10^{-5},\; 5.6 \times 10^{-8}$), ($3.4\times 10^{-4},\; 4.2 \times 10^{-7}$), and ($2.1\times 10^{-3},\; 2.6 \times 10^{-5}$), respectively. }  
\label{fig:arxiv100}
\end{figure}

We also experiment with an ensemble of 100 model realizations to understand the sensitivity associated with the model outputs. Figure~\ref{fig:arxiv100} shows degree distribution of real-network ArXiv-HepPH compared against \shortname{}. Small standard deviations over $100$ model realizations demonstrate low variability in the generated outputs. Similar observations are obtained for CP, CPT, and FF. CP, CPT, and FF with standard deviations for 100  model realizations are $9.9\times 10^{-5}$, $3.4\times 10^{-4}$, and $2.1\times 10^{-3}$, respectively. Similar results were obtained for other real-networks.

\begin{table*}
\def\bcell{\cellcolor{gray80}}
\centering
\resizebox{\hsize}{!}{
\begin{tabular}{|c|l| c|c|c|c|c|c|} \hline
Statistic&Networks & Observed & CP & CPT & FF & \shortname1 & \shortname2   \\ \hline
\multirow{4}{*}{\rotatebox[origin=c]{90}{Triangles}}&Biomedical& $6.2\times10^6$ &  $0.45$ & \cellcolor{red!15} 0.10 & $0.60$ & \cellcolor{green!20}$\textbf{1.05}$& $0.45$  \\
&Supreme court & $7.1\times 10^6$ &$0.21$& \cellcolor{red!6} 0.24  &$0.59$  & $0.73$ &\cellcolor{green!20}$\textbf{0.98}$  \\
&ArXiv-HepTH & $3.4\times 10^7$  &$0.09$& \cellcolor{red!15} 0.35 & $1.94$  & $0.68$&\cellcolor{green!20}$ \textbf{0.76}$\\
&ArXiv-HepPH & $2.3\times 10^7$ & $ 0.09$& \cellcolor{red!6} 0.56 & $0.83$ & $0.48$&\cellcolor{green!20}$\textbf{0.96}$\\\hline
\multirow{4}{*}{\rotatebox[origin=c]{90}{Diameter}}&Biomedical &57.8 & 0.51 &\textbf{1.3} & 0.36 &0.57 &0.56 \\
&Supreme court& 10.3 & 1.14 &3.0 & \textbf{0.87} &1.45 & 1.7\\
&ArXiv-HepTH & 17.8& 0.59 &0.66 & 0.43 &\cellcolor{green!15}\textbf{0.92} & \cellcolor{green!8} 1.22 \\
&ArXiv-HepPH & 15.0& 0.81 &0.82 & 0.74 &1.23 & \cellcolor{green!15} \textbf{1.01} \\
\hline
\multirow{4}{*}{\rotatebox[origin=c]{90}{H-index}}&Biomedical &84 & 82 &20 & 81 &94 &\cellcolor{green!15}\textbf{84} \\
&Supreme court& 89 & 85 &39 & \textbf{87} &94 &\cellcolor{green!15} \textbf{87}\\
&ArXiv-HepTH & 170& 115 &61 & 193 &\cellcolor{green!8}155 & \cellcolor{green!15} \textbf{175} \\
&ArXiv-HepPH & 158& 125 &67 & 143 &175 & \cellcolor{green!15} \textbf{160} \\
\hline
\end{tabular}}
\caption{Real and simulated triangle counts, average diameter and h-index over the lifetime of four networks. The third column (row 2--5) shows the number of triangles observed in each data network. Subsequent columns show the ratio between the simulated and observed numbers of triangles. A ratio close to 1 indicates a better model driving the simulation.  \shortname1 and \shortname2 generally achieve the ratios closest to~1. The third column (row 6--9) shows average diameter for the real data set. Subsequent columns show the ratio between the simulated and observed average diameters. A ratio close to 1 indicates a better model. Similarly, the third column (row 10--13) shows h-index of real networks compared against simulated h-index (column 4--8).
}  \label{t2}
\end{table*}

\begin{table*}
\def\bcell{\cellcolor{gray80}}
\centering
\resizebox{\hsize}{!}{
\begin{tabular}{|c|l|c|c|c|c|c|} \hline
Statistic&Metric  & CP & CPT & FF & \shortname1 & \shortname2   \\ \hline
\multirow{2}{*}{\rotatebox[origin=c]{0}{Triangles}}&Mean&   $0.14$ &0.59 & $0.69$ & 1.05& $\textbf{0.97}$  \\
&STD &$0.02$&  0.002  &$0.24$  & $0.22$ &0.19  \\\hline
\multirow{2}{*}{\rotatebox[origin=c]{0}{Diameter}}&Mean & 0.79 &0.83 & 0.76 &1.18 &\textbf{1.09} \\
&STD& 0.05 &0.03 & 0.09&0.10 & 0.08\\\hline
\multirow{2}{*}{\rotatebox[origin=c]{0}{H-index}}&Mean &0.49 &0.43 & 0.78 &\textbf{1.03} &0.96 \\
&STD& 0.012 &0.013 & 0.122 &0.113 &0.115\\\hline
\end{tabular}}
\caption{Mean values of the ratio of simulated to real triangle counts, average diameter and h-index over the lifetime of ArXiv-HepPH network is reported in the table along with standard deviation (STD). A ratio close to 1 indicates a better model driving the simulation. Our proposed models are performing well as compared to CP, CPT, and FF. (Mean$\pm$STD) means $68\%$ models networks would have ration between (Mean$-$STD) and (Mean$+$STD), (Mean$\pm$2STD) means $95\%$ models networks would have ration between (Mean$-$STD) and (Mean$+$STD), and (Mean$\pm$STD) means $99.7\%$ models networks would have ration between (Mean$-$STD) and (Mean$+$STD). From the discussion, it is observed that in case of CP and CPT it is very rare to get ratio 1, because mean and standard deviation both are very less that results: For CP,  (Mean$\pm$3STD) in to (0.08 to 0.2), (0.6 to 0.84), and (0.454 to 0.506) for triangles, average diameter, and h-index, respectively, For CPT, (Mean$\pm$3STD) in to (0.584 to 0.596), (0.77 to 0.89), and (0.391 to 0.469) for triangles, average diameter, and h-index, respectively. FF has $68\%$ model networks which have ratio for triangles, average diameter, and h-index in the ranges (0.45 to 0.93), (0.67 to 0.85), and (0.658 to .902), respectively. \shortname1 has $68\%$ model networks which have ratio for triangles, average diameter, and h-index in the ranges (0.83 to 1.27), (1.08 to 1.28), and (0.917 to 1.13), respectively. \shortname2 has $68\%$ model networks which have ratio for triangles, average diameter, and h-index in the ranges (0.78 to 1.16), (1.01 to 1.17), and (0.845 to 1.075), respectively.}  \label{tab:sensitivity}
\end{table*}

\subsection{Triangle counts}
In Table~\ref{t2} we report the number of triangles present in the real datasets. In addition, we also report the ratio between the number of triangles obtained from the simulated networks (CP model, CPT model, FF model, \shortname1 and \shortname2 models) and the real networks for each of the datasets. We observe that our models match the real data much better than all the other three models. Table~\ref{tab:sensitivity} shows mean and standard deviation of ratio  of simulated and real triangle counts for 100 realizations of above models. As expected, we observe significantly low standard deviation values.

\subsection{Average diameter}
Table \ref{t2} reports the average diameter over the lifetime for various real networks in the second column (Observed), where the step size is 5000 in terms of the number of nodes. Similarly, for all the competing models, we compute the diameter of the network at each step, and take an average. The ratio of the average value of the diameter obtained by the model and observed value is shown in the table. In two of the four data sets, \shortname{} variants provide simulated diameters closest to the observed diameters. Note that the CPT model has advantage over the other models because it uses degree sequence from the real dataset. Table~\ref{tab:sensitivity} shows mean and standard deviation of ratio  of simulated and real average diameter for 100 realizations of above models. As expected, we observe significantly low standard deviation values.

\subsection{H-index}
In Table~\ref{t2}, we report h-index of the real datasets. Additionally, we also compute h-index of the networks obtained under different network models considered in this paper (CP model, CPT model, FF model, \shortname1 and \shortname2 models). We observe that our models match the real data much better than all the other three models. This result indicates that our model is able to much better replicate the h-index of the network and can therefore find an application in predicting the h-index of authors and journals in future.  Table~\ref{tab:sensitivity} shows mean and standard deviation of of ratio  of simulated and real h-index for 100 realizations of above models. As expected, we observe significantly low standard deviation values.

\subsection{Obsolescence}

It is well known that the number of citations to a randomly sampled  article does not keep growing over time \citep{Singh:2017:RMP:3097983.3098146,waumans2016genealogical,Wang20094273,wang2013quantifying}.  The rate of acquisition of citations is known to rise to a peak between three and five years (for most communities) and then decline, sometimes sharply. 
Nevertheless, PA, CP and related models favor the growth of citations to older nodes; age is always an asset and never a liability.  This sharply contradicts observed data, where a vast majority of old papers are eventually forgotten.  Thus, PA-style models overestimate the popularity of old papers and underestimate the popularity of younger papers.

Singh et al.~\cite{Singh:2017:RMP:3097983.3098146} proposed a temporal sketch of a network's evolution history that captures obsolescence dynamics.  We adapt it slightly for our use here.  Consider the $o$\% oldest nodes, and count their total degree at the end of time.  Divide by the total degree over all nodes at the end of time.  This ratio $r$ grows with $o$ to a maximum of~1 when 100\% of the nodes are included.  The more quickly $r$ grows with $o$, the closer the situation is to PA.  In contrast, slower growth of $r$ with increasing $o$ indicates a strong effect of obsolescence.

\begin{figure*}[!tbh]
\def\halfwidth{.47\textwidth}
\centering
\begin{tabular}{cc}
\includegraphics[width=0.4\textwidth]{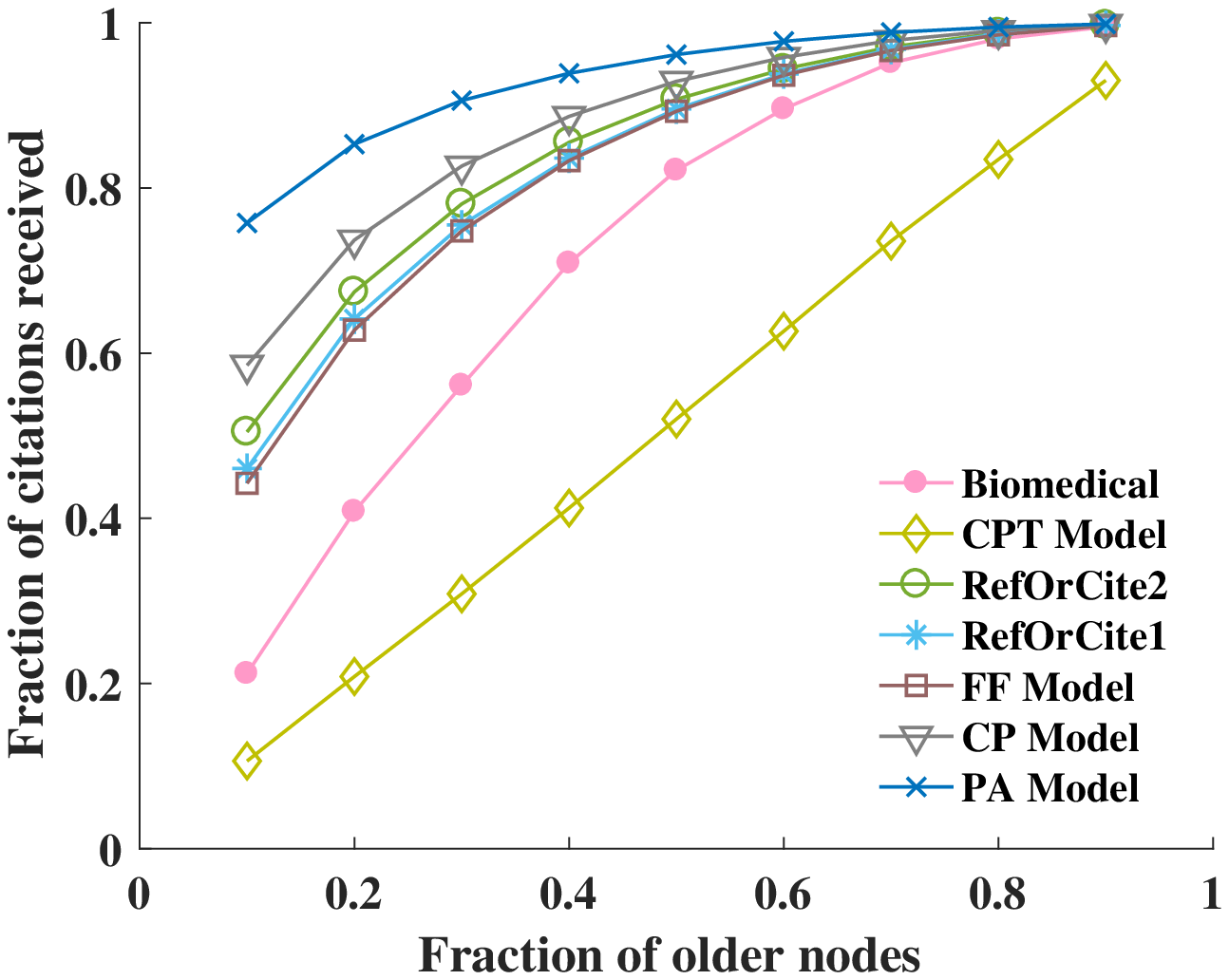} &
\includegraphics[width=0.4\textwidth]{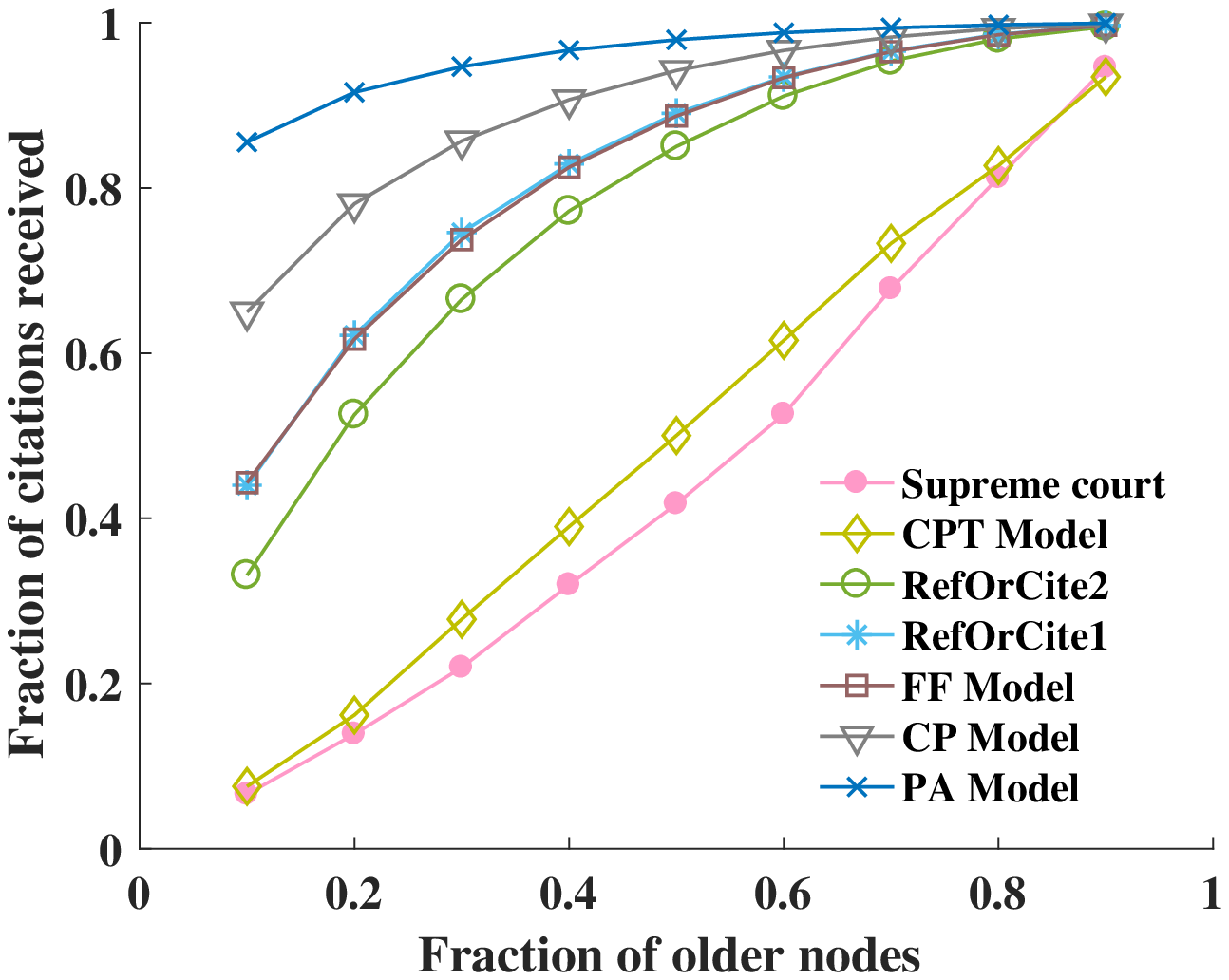}\\
(a) Biomedical  & (b) Supreme court\\
\includegraphics[width=0.4\textwidth]{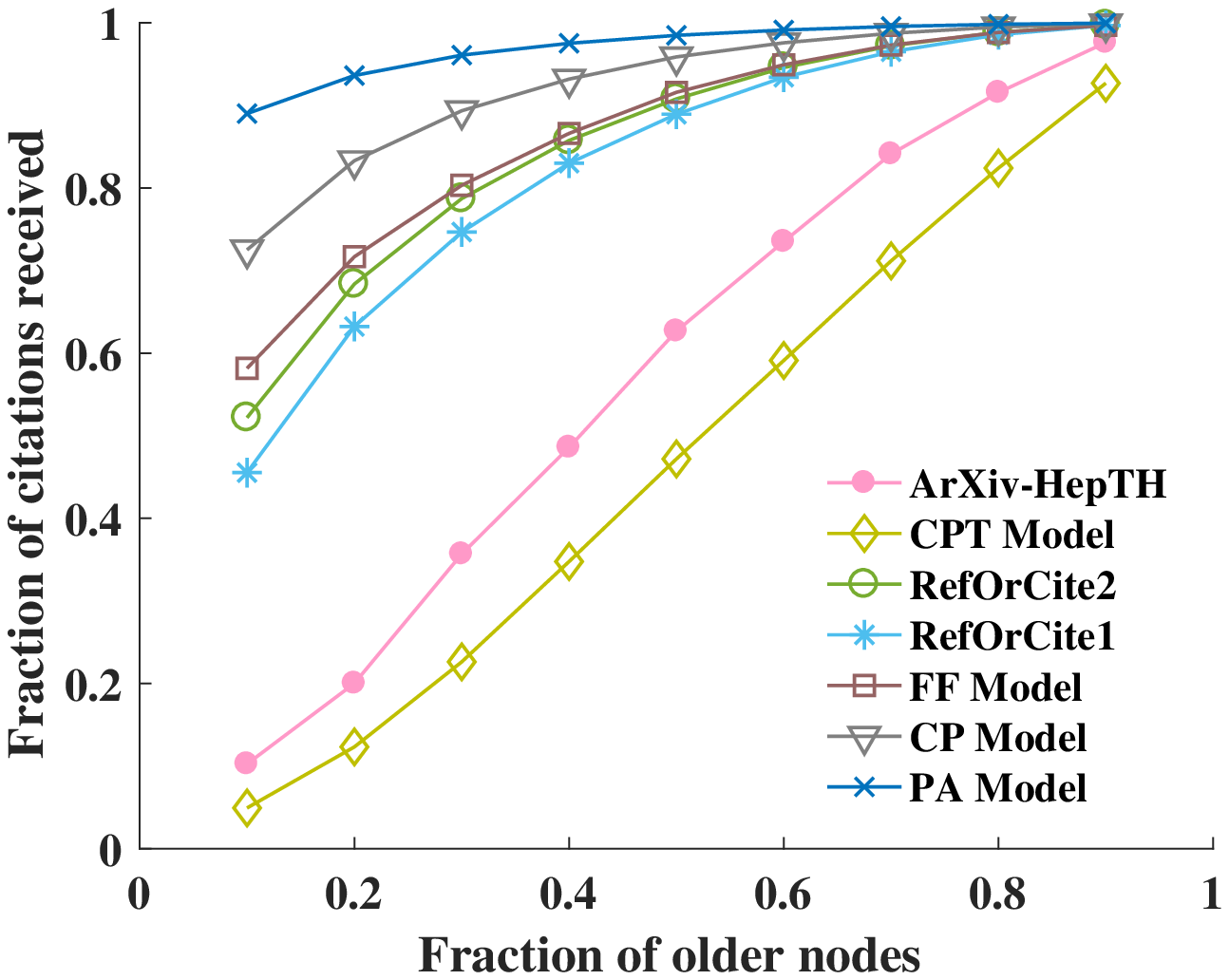} &
\includegraphics[width=0.4\textwidth]{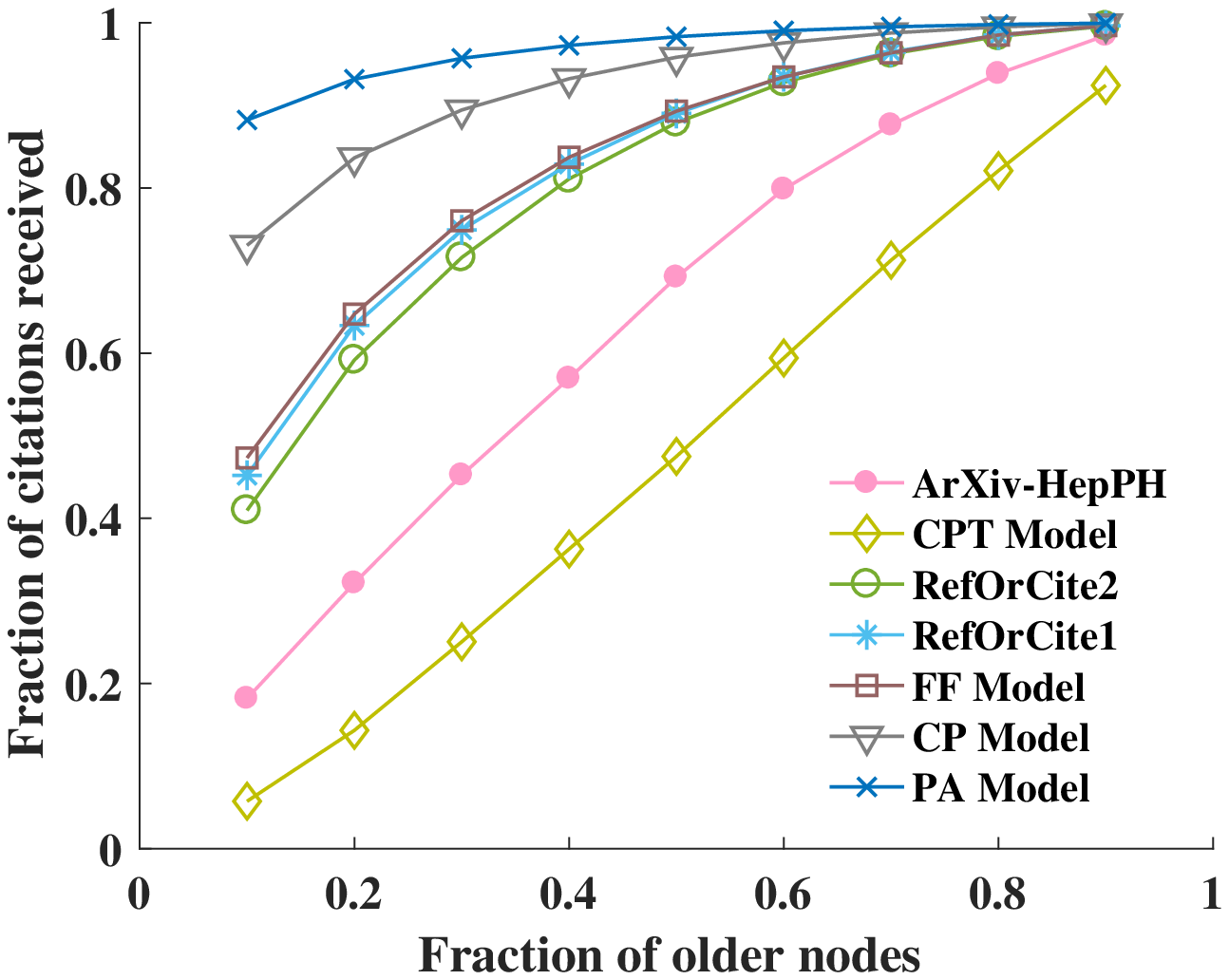}\\
(c) ArXiv-HepTH & (d) ArXiv-HepPH\\
\end{tabular}
\caption{(Best viewed in color.)  Fraction of citations received by the oldest $o$ fraction of nodes: (a)~Biomedical, (b)~Supreme court, (c)~ArXiv-HepTH, and (d)~ArXiv-HepPH.  A comparison among real data (pink), CPT model (yellow), CPT model (gray), BA model (blue), \shortname1 (cyan), \shortname2 model (green) and FF (violet).}  \label{fig:relay1}
\end{figure*}

\figurename~\ref{fig:relay1} shows $r$-against-$o$ plots for different data sets.  For each data set, the real network shows one trajectory.  A model is faithful to the obsolescence behavior of the real network if its trajectory is close to the real trajectory.  With the exception of CPT, the models closest to the real trajectory are \shortname1 and \shortname2.  By allowing links from new nodes to (more recent) inlinks of the base node, they naturally model obsolescence.  In contrast, PA and CP, as expected, confer undue popularity to older nodes (very large $r$ for small $o$).  Curiously, \shortname{} is as good as FF in most cases.  Although CPT models obsolescence better than \shortname, since it is the only model that incorporates a link  probability that depends on paper age, its match with degree distribution and triangle count are far worse than \shortname. This under performance of the CPT model can be attributed to the fact that this model is most suitable for networks where the node degrees grow very slowly over time \cite{krapivsky2005network}, as opposed to the data sets we study, where the node degrees increase relatively fast.

We do not compare our model with other mechanistic growth models \cite{Singh:2017:RMP:3097983.3098146,Wang20094273} that are primarily composed of PA with an age-based decay component (often exponential), because this mechanism inherently limits triangle formation.  \cite{Wang20094273} multiply PA's linking probability with an age-based exponential decay term.  It suffers from similar limitations of clustering and triangle formation as PA.  The exponential decay factor only restricts the growth of degrees of the nodes to incorporate aging.

\section{Conclusion and future work}
\label{sec:end}

Idealized network evolution models that explain preferential attachment in citation networks are abundant, but only a few analyze citation and reference copying.  We present \shortname: novel network-driven models to explain triangle formation and obsolescence in real bibliographic networks.  We conduct formal analysis of various properties of \shortname{} to establish behavior expected from real networks. Traditional growth models do not fit the real data well, but our \shortname{} models do.  Overall, \shortname{} fits the largest number of important network properties better than other proposals.

However, a number of potential limitations remain to be addressed. First, the current study employs relatively small bibliographic datasets.  Therefore, we do not claim generic applicability on very large bibliographic networks.  In future, we plan to extend this study to other citation networks, for example, patent citation networks.  Second, our proposed \shortname{} models do not consider topic or author (collaboration) information which might be relevant in copying citations and references.

\section{References}
\bibliographystyle{elsarticle-harv}
\bibliography{ref,apssamp}
\end{document}